# Eigenvalues of Ising Hamiltonian with long-range interactions

L.B. Litinskii and B.V. Kryzhanovsky

Center of Optical Neural Technologies,
Scientific Research Institute for System Analysis of Russian Academy of Sciences,
Nakhimov ave, 36-1, Moscow, 117218, Russia
*litin@mail.ru   kryzhanov@mail.ru*

**Abstract**

We obtained analytically eigenvalues of a multidimensional Ising Hamiltonian on a hypercube lattice and expressed them in terms of spin-spin interaction constants and the eigenvalues of the one-dimensional Ising Hamiltonian (the latter are well known). To do this we wrote down the multidimensional Hamiltonian eigenvectors as the Kronecker products of the eigenvectors of the one-dimensional Ising Hamiltonian. For periodic boundary conditions, it is possible to obtain exact results taking into account interactions with an unlimited number of neighboring spins. In this paper, we present exact expressions for the eigenvalues for the planar and cubic Ising systems accounting for the first five coordination spheres (that is interactions with the nearest neighbors, the next neighbors, the next-next neighbors, the next-next-next neighbors and the next-next-next-next neighbors). In the case of free-boundary systems, we showed that in the two- and three-dimensions the exact expressions could be obtained only if we account for interactions with spins of first two coordination spheres and first three coordination spheres, respectively.

## I. Introduction

The Ising model remains in a focus of scientific interest for decades. Partially it is because this model is convenient when testing new methods of calculation of the free energy and critical characteristics. In the same time, in spite of a seemingly straightforward setting of the problem, in some cases the scientists could not find its rigorous solutions. In paper [1], one can find a review of different approaches to the solution of the Ising problem and the obtained results. There is a number of books, where different applications of the model to analysis of different physical problems are collected. Namely, in the book [2] phase transitions in solids are discussed in terms of the Ising model; the model's applications to spin glasses and the neural network theory can be found in the monograph [3]; and the collective monograph [4] describes the link between the Ising model and optimization problems.

In the present paper, we obtain exact expressions for the eigenvalues of the Ising Hamiltonian on the hypercube lattices taking into account an arbitrary number of neighbors and analyze the dependence of the obtained results on the type of boundary conditions. Our results generalize the analysis presented in [5].

Usually, when characterizing a proximity between a given and other spins it is common to use the expressions "the nearest neighbor", "the next neighbor", "the next-next neighbor". However, since we consider interactions with spins that are far enough from the given spin, in place of repeating the "next" prefixes we use the concept of coordination spheres [6, 7]. Let us organize an increasing sequence of various distances from a given spin to all other spins. Then a $k$-th coordination sphere includes all the nodes whose distance to the given node takes the $k$-th place in this sequence. Clearly, all the spins belonging to a coordination sphere interact equally with the given spin. Let $w_k$ be the interaction constant with the spins belonging to the $k$-th coordination sphere. When the value of $k$ is not too large, the common notations are convenient. Namely, the nearest neighbors belong to the first coordination sphere, the next neighbors to the second coordination sphere, the next-next neighbors to the third coordination sphere. In what follows we use the coordination sphere numbers as well as the common notations.

In Section II, we discuss the one-dimensional Ising model in detail. We introduce matrices $\mathbf{J}(k)$ that describe the interactions with the spins of the $k$-th coordination sphere. In the case of periodic boundary conditions, all the matrices $\mathbf{J}(k)$ commute. This allows us to obtain easily an expression for the eigenvalues of the one-dimensional Hamiltonian $\mathbf{H}_1 = \sum_k w_k \mathbf{J}(k)$, where we take into account an arbitrary number of the coordination spheres. On the contrary, if we assume free boundary conditions the matrices $\tilde{\mathbf{J}}(k)$ do not commute. (In what follows we use the tilde symbol to denote characteristics related to free boundary conditions.) As a result, it is not possible to express the eigenvalues of the matrix $\tilde{\mathbf{H}}_1 = \sum_k w_k \tilde{\mathbf{J}}(k)$ in terms of the eigenvalues of the individual matrices $\tilde{\mathbf{J}}(k)$. In this case, we know only the eigenvalues $\{\tilde{\lambda}_i\}_1^n$ of the matrix $\tilde{\mathbf{J}}(1)$ that accounts for the interaction $w_1$ with the nearest neighbors.

In Section III, we examine the two- and three-dimensional Ising models with periodic boundary conditions. In the two-dimensional case, we account for the interactions with the spins of the first and second coordination spheres. In the three-dimensional case, we add the interaction with the spins of the third coordination sphere. We restrict our consideration to these relatively simple models due to the following reasons. Firstly, here we present in detail our method, which allows us to define the eigenvalues of the multidimensional Hamiltonian in terms of the interaction constants and the eigenvalues of the one-dimensional Hamiltonian. This method is rather cumbersome, so it is convenient to use simple examples. Secondly, as we show in Section IV, the obtained results are fully applicable also in the case of free boundaries. In this Section, we also show that for free boundary conditions it is impossible to extend these results to the case of more neighbors.

Finally, in Section V we develop one other method for calculation of the eigenvalues of the multidimensional Ising Hamiltonian for periodic boundary conditions. We base our calculation on representing the eigenvectors of the multidimensional problem as the Kronecker product of the eigenvalues of the one-dimensional Ising Hamiltonian. The method is especially effective when we account for a large number of coordination spheres. As an example, for the planar and cubic lattices we obtain exact expressions taking into account five coordination spheres.

The obtained results can be useful when analyzing spectral densities of the multidimensional Ising systems and their dependencies on the parameters of the spin-spin interactions. It is possible that our exact expressions for the eigenvalues can be applied when studying optical transitions and in calculations of the free energy of spin systems. In addition, they can be significant in the analysis of the role of long-range hopping in many-body localization for lattice systems of various dimensions (see [8] and references therein). For natural spin systems, the interaction constants are typically determined by the distances between the spins. Then truncating the number of interactions by accounting only for a finite number of coordination spheres is an approximation, which holds the better the stronger the coordinate dependence of the interaction. However, for artificial spin systems with couplers, such as the ones used for quantum annealing (see for example [9]), our results with a finite number of coordination spheres can be viewed as exact.

## II. 1D-Ising model

In this Section, we examine the one-dimensional Ising model that describes a linear chain of $n$ connected spins $s_i = \pm 1$, $i = 1,..,n$. In Subsection 1, we assume periodic boundary conditions. This implies that the chain of spins forms a closed loop: the $n$-th spin of the chain is said to be the left nearest neighbor of the first spin, the second last $(n-1)$-th spin turns out to be the next left neighbor of the first spin and so on. In this case we can obtain some rigorous results. In Subsection 2, we turn to free boundary conditions when there are no additional conditions imposed on the last spins of the chain. This is a more complex problem for analytical calculations.

**1. Periodic boundary conditions.** Let $\mathbf{J}(k)$ be a connection matrix of the one-dimensional Ising system where we account for the interaction of each spin with its $k$-th neighbor only. For example, the matrices $\mathbf{J}(1)$ and $\mathbf{J}(2)$ have the form

$$\mathbf{J}(1) = \begin{pmatrix} 0 & 1 & 0 & 0 & . & . & 1 \\ 1 & 0 & 1 & 0 & . & . & 0 \\ 0 & 1 & 0 & 1 & . & . & 0 \\ . & . & . & . & . & . & . \\ 0 & 0 & . & . & 1 & 0 & 1 \\ 1 & 0 & 0 & . & . & 1 & 0 \end{pmatrix}, \quad \mathbf{J}(2) = \begin{pmatrix} 0 & 0 & 1 & 0 & . & 1 & 0 \\ 0 & 0 & 0 & 1 & . & . & 1 \\ 1 & 0 & 0 & 0 & . & . & 0 \\ . & . & . & . & . & . & 1 \\ 1 & 0 & . & 1 & 0 & 0 & 0 \\ 0 & 1 & 0 & . & 1 & 0 & 0 \end{pmatrix}.$$

It is evident that $\mathbf{J}(k)$ is a symmetric matrix where the ones occupy the $k$-th and $(n-k)$-th superdiagonals parallel to the matrix main diagonal. Since $\mathbf{J}(k) \equiv \mathbf{J}(n-k)$, we restrict the domain of $k$ to $k \leq k_0 = \left[\dfrac{n}{2}\right]$. Matrices $\{\mathbf{J}(k)\}_1^{k_0}$ allows us to write easily the one-dimensional Hamiltonian when we account for the interactions with the spins of the first coordination sphere $w_1$, the second coordination sphere $w_2$, ... and the $k$-th coordination sphere $w_k$:

$$\mathbf{H}_1 = w_1 \cdot \mathbf{J}(1) + w_2 \cdot \mathbf{J}(2) + ... + w_k \cdot \mathbf{J}(k). \tag{1}$$

Now let us calculate the eigenvalues of the matrix $\mathbf{J}(1)$. The relevant system of equations has the form

$$\begin{array}{rcl}
0 + x_2 + 0 \ldots + x_n &=& \lambda \cdot x_1 \\
x_1 + 0 + x_3 \ldots + 0 &=& \lambda \cdot x_2 \\
\ldots \ldots \ldots \ldots \ldots \ldots \ldots \ldots \ldots \\
x_1 + 0 + 0 \ldots x_{n-1} + 0 &=& \lambda \cdot x_n
\end{array} \tag{2}$$

The solution of this system of equations is well-known (see [5] and [10]):

$$\lambda_i(1) = 2 \cdot \cos(\varphi_i), \text{ where } \varphi_i = \dfrac{2\pi \cdot (i-1)}{n}, \quad i = 1, 2, ..., n. \tag{3}$$

We also know the eigenvectors corresponding to the system (2) (see [5] and [11]). Since almost all the eigenvalues (3) are twice degenerate, there are different ways to write down these eigenvectors. Their most convenient form is

$$\mathbf{J}(1) \cdot \mathbf{f}_i = \lambda_i(1) \cdot \mathbf{f}_i, \quad \mathbf{f}_i = \begin{pmatrix} f_i^{(1)} \\ f_i^{(2)} \\ \vdots \\ f_i^{(n)} \end{pmatrix}, \text{ where } f_i^{(j)} = \dfrac{\cos(j-1)\varphi_i + \sin(j-1)\varphi_i}{\sqrt{n}}, \quad i, j = 1, 2, ..., n. \tag{4}$$

Now let us calculate the eigenvalues of the matrix $\mathbf{J}(k)$ when $k > 1$. When raising subsequently the matrix $\mathbf{J}(1)$ to integer powers, we obtain the following relations

$$\begin{aligned}
\mathbf{J}(2) &= \mathbf{J}^2(1) - 2 \cdot \mathbf{I}, \\
\mathbf{J}(3) &= \mathbf{J}^3(1) - 3 \cdot \mathbf{J}(1), \\
\mathbf{J}(4) &= \mathbf{J}^4(1) - 4 \cdot \mathbf{J}^2(1) + 2 \cdot \mathbf{I}, \\
&\ldots\ldots\ldots
\end{aligned} \tag{5}$$

Here $\mathbf{I} = \mathbf{diag}(1, 1, ..., 1)$ is a diagonal $(n \times n)$ identity matrix. Now we will show that for any $k$ the matrix $\mathbf{J}(k)$ is a polynomial of the matrix $\mathbf{J}(1)$. Let us use an equality which is easy to verify:

$$\mathbf{J}(k) \cdot \mathbf{J}(l) = \begin{cases} \mathbf{J}(k+l) + \mathbf{J}(k-l), & \text{when } k > l \\ \mathbf{J}(2k) + 2 \cdot \mathbf{I}, & \text{when } k = l \end{cases}. \tag{6}$$

Suppose we have shown that the matrix $\mathbf{J}(l)$ is a polynomial of the matrix $\mathbf{J}(1)$ for all $l \leq k$. Then from Eq. (6) it follows that for any value of $k + l$ we can also present the matrix $\mathbf{J}(k+l)$ as a polynomial of $\mathbf{J}(1)$. Then for any $k$ the matrix $\mathbf{J}(k)$ is a polynomial of the matrix $\mathbf{J}(1)$, and consequently all the matrices $\mathbf{J}(k)$ commute. This means that all these matrices have the same set of eigenvectors (4).

Now let us obtain the expressions for the eigenvalues $\{\lambda_i(k)\}_{i=1}^n$ of the matrix $\mathbf{J}(k)$. From the first of the equations (5) it follows that

$$\lambda_i(2) = \lambda_i^2(1) - 2 = 2\left(2 \cdot \cos^2(\varphi_i) - 1\right) = 2 \cdot \cos(2 \cdot \varphi_i), \quad i = 1, 2, ...n.$$

Likewise, when we use the second of the equations (5), we obtain $\lambda_i(3) = 2 \cdot \cos(3 \cdot \varphi_i)$. After that with the aid of the equality (6), we easily prove that for any $k$

$$\lambda_i(k) = 2 \cdot \cos(k \cdot \varphi_i), \quad i = 1, 2, ..., n. \tag{7}$$

Indeed, let the equality (7) be fulfilled for all $l \leq k$. Then with the aid of Eq. (6) we obtain

$$\lambda_i(k+l) = 2\big(2\cos(k\varphi_i) \cdot \cos(l\varphi_i) - \cos((k-l)\varphi_i)\big) = ... = 2 \cdot \cos((k+l)\varphi_i).$$

We note that since all the matrices $\mathbf{J}(k)$ have the same set of the eigenvectors $\{\mathbf{f}_i\}_1^n$, we can obtain an explicit form of the eigenvalues of the one-dimensional Hamiltonian $\mathbf{H}_1$ (1):

$$\lambda_i(\mathbf{H}_1) = w_1 \cdot \lambda_i(1) + ... + w_k \cdot \lambda_i(k) = 2\sum_{l=1}^{k} w_l \cdot \cos(l \cdot \varphi_i), \text{ when } i = 1, 2, ..., n. \tag{8}$$

**2. Free boundaries.** We recall that in this case we use the tilde symbol to define the same characteristics as in Subsection 1. Now when composing the matrix $\tilde{\mathbf{J}}(k)$, which is an analog of the matrix $\mathbf{J}(k)$, again we take into account the interaction of a given spin with its $k$-th neighbors. However, in contrast to the case of periodic boundary conditions, not all the spins are equivalent, and the spins that are closer to the ends of the chain may not have one of two possible neighbors. The matrix $\tilde{\mathbf{J}}(k)$ is symmetric; its $k$-th superdiagonal is nonzero (as well as the symmetric subdiagonal). However, contrary to the case of periodic boundary conditions, the elements of the $(n-k)$-th superdiagonal are equal to zero.

We know the expressions for the eigenvalues and eigenvectors of the matrix $\tilde{\mathbf{J}}(1)$ [12], [13]:

$$\tilde{\lambda}_i(1) = 2 \cdot \cos(\tilde{\varphi}_i), \text{ where } \tilde{\varphi}_i = \frac{\pi \cdot i}{n+1}, \text{ and } \tilde{\mathbf{f}}_i = \begin{pmatrix} \tilde{f}_i^{(1)} \\ \tilde{f}_i^{(2)} \\ \vdots \\ \tilde{f}_i^{(n)} \end{pmatrix}, \tilde{f}_i^{(j)} = \sqrt{\frac{2}{n+1}} \sin(j \cdot \tilde{\varphi}_i), i, j = 1, 2, ..., n. \tag{9}$$

The equations (9) as well as the equations (3) and (4) are easy to obtain by simple trigonometric calculations.

However, as it can be easily verified, the matrices $\tilde{\mathbf{J}}(k)$ do not commute. This means that each matrix $\tilde{\mathbf{J}}(k)$ has its own set of eigenvectors and all these sets are different. Consequently, we cannot define a single basis allowing us to diagonalize all the matrices $\tilde{\mathbf{J}}(k)$ simultaneously and to obtain the equations of the form (8). We will show that this is the reason why it is much more difficult to obtain rigorous results for the two- and three-dimensional Ising systems with the free boundaries. More specifically, for periodic boundary conditions we can express the eigenvalues of the multidimensional Hamiltonian in terms of the eigenvalues of the one-dimensional Hamiltonian $\{\lambda_i(1)\}_1^n$ taking into account interactions with the spins of an arbitrary number of the coordination spheres. In general, for systems with free boundaries such possibility does not exist. We will discuss the exceptions in the end of the next Section.

**Note.** In general, in each row (and each column) of the matrices $\mathbf{J}(k)$ there are two ones. However, when $n = 2m$ and $k = m$, the lower and the upper superdiagonals coincide: $k = m = n - k$. Such a matrix has only a single one in each its row. For example, when $n = 6, k = 3$ we have

$$\mathbf{J}(k=3) = \begin{pmatrix} 0 & 0 & 0 & 1 & 0 & 0 \\ 0 & 0 & 0 & 0 & 1 & 0 \\ 0 & 0 & 0 & 0 & 0 & 1 \\ 1 & 0 & 0 & 0 & 0 & 0 \\ 0 & 1 & 0 & 0 & 0 & 0 \\ 0 & 0 & 1 & 0 & 0 & 0 \end{pmatrix}.$$

Here we have a contradiction: from Eq. (7) it follows that $\lambda_i(3) = 2\cos(\pi \cdot (i-1)) = \pm 2$, $i = 1, 2, ..n$, but a direct calculation gives the values that are two times less: $\lambda_i(3) = \pm 1$.

When $n$ is sufficiently large, the matrix $\mathbf{J}(n/2)$ practically does not enter the obtained results (see Sections III and IV). However, generally speaking, this contradiction has to be resolved. For this purpose, we can change slightly the generation algorithm for the matrices $\mathbf{J}(k)$. Namely, instead of writing the ones along the $k$-th and $(n-k)$-th superdiagonals, we have to add the ones to the matrix elements along these superdiagonals. For the matrices in question, this change leads to appearance of the twos in places of the ones, but it does not involve all the other matrices.

### III. 2D- and 3D- Ising systems with periodic boundary conditions: simple case

In this Section, we analyze rather simple models of the two-dimensional and three-dimensional Ising systems with periodic boundary conditions. In the case of two dimensions, we take into account the interactions with the spins from the first and the second coordination spheres (Subsection 1). When considering the three-dimensional model, we account for the first, the second, and the third coordination spheres (Subsection 2).

Let us agree on notations. For the lattices of both dimensions, the spins of the first coordination sphere are at a distance equal to the lattice constant from the given spin. Let this distance be equal to 1 and let $w_1$ be the corresponding interaction constant. Next, the spins at the second coordination sphere are at a distance equal to $\sqrt{2}$ from the given spin (see Fig. 1) and $w_2$ is the constant of interaction with these spins. In the case of the cubic lattice, we will also need to examine spins from the third coordination sphere, whose distance from the given spin is equal to $\sqrt{3}$ and the corresponding interaction constant is $w_3$.

The symbols $\mathbf{A}_1$, $\mathbf{B}_1$, $\mathbf{C}_1$ ... denote $(n \times n)$ matrices that, in analogy with Eq.(1), are the weighted sums of the matrices $\mathbf{J}(k)$. The subscript "1" indicates that these matrices relate to the one-dimensional Ising system. In two dimensions, to describe the interactions we need $(n^2 \times n^2)$ matrices $\mathbf{A}_2$, $\mathbf{B}_2$, $\mathbf{C}_2$, where again the subscript "2" indicates the dimension of the problem. Finally, a $(n^3 \times n^3)$ matrix $\mathbf{A}_3$ describes the interactions in the cubic Ising system (for details see Subsection 2).

Finally, to simplify the obtained expressions from here and to the end of the paper we use the notation $\lambda_i$ in place of $\lambda_i(1)$ for the eigenvalues of the matrix $\mathbf{J}(1)$.

**1.** In the two-dimensional square Ising system, the total number of spins is equal to $n^2$, where $n$ is the number of spins at the side of the square. By $\mathbf{A}_2$ we denote the corresponding $(n^2 \times n^2)$- connection matrix. The explicit form of its first $n$ rows is

$$\begin{pmatrix} 0 & w_1 & 0 & \cdots & w_1 & w_1 & w_2 & 0 & \cdots & w_2 & 0 & \cdots & 0 & \cdots & w_1 & w_2 & 0 & \cdots & w_2 \\ w_1 & 0 & w_1 & \cdots & 0 & w_2 & w_1 & w_2 & \cdots & 0 & 0 & \cdots & 0 & \cdots & w_2 & w_1 & w_2 & \cdots & 0 \\ 0 & w_1 & 0 & w_1 & 0 & 0 & w_2 & w_1 & w_2 & 0 & 0 & \cdots & 0 & \cdots & 0 & w_2 & w_1 & w_2 & 0 \\ \cdots & \cdots & \cdots & \cdots & \cdots & \cdots & \cdots & \cdots & \cdots & \cdots & 0 & \cdots & 0 & \cdots & \cdots & \cdots & \cdots & \cdots & \cdots \\ w_1 & 0 & 0... & w_1 & 0 & w_2 & 0 & 0... & w_2 & w_1 & 0 & \cdots & 0 & \cdots & w_2 & 0 & 0... & w_2 & w_1 \end{pmatrix}. \qquad (10)$$

$$\underbrace{\phantom{xxxxxx}}_{n} \quad \underbrace{\phantom{xxxxxx}}_{n} \quad \underbrace{\phantom{xxxxxx}}_{n}$$

Rewriting Eq. (10) in a block form, we obtain

$$\left( w_1 \mathbf{J}(1) \quad w_1 \mathbf{I} + w_2 \mathbf{J}(1) \quad \mathbf{0}....\mathbf{0} \quad w_1 \mathbf{I} + w_2 \mathbf{J}(1) \right),$$

where $\mathbf{J}(1)$ and the identity matrix $\mathbf{I}$ was introduced in the previous Section. Let us define $(n \times n)$ matrices

$$\mathbf{A}_1 = w_1 \mathbf{J}(1) \text{ and } \mathbf{B}_1 = w_1 \mathbf{I} + w_2 \mathbf{J}(1) \qquad (11)$$

and present the whole interaction matrix $\mathbf{A}_2$ in the block form:

$$\mathbf{A}_2 = \begin{pmatrix} \mathbf{A}_1 & \mathbf{B}_1 & 0 & 0 & \cdots & \mathbf{B}_1 \\ \mathbf{B}_1 & \mathbf{A}_1 & \mathbf{B}_1 & 0 & \cdots & 0 \\ 0 & \mathbf{B}_1 & \mathbf{A}_1 & \mathbf{B}_1 & \cdots & 0 \\ \cdots & \cdots & \cdots & \cdots & \cdots & \cdots \\ \mathbf{B}_1 & 0 & \cdots & 0 & \mathbf{B}_1 & \mathbf{A}_1 \end{pmatrix}. \qquad (12)$$

When generating the matrix $\mathbf{A}_2$ we use two $(n \times n)$-matrix blocks $\mathbf{A}_1$ and $\mathbf{B}_1$ defined in Eq. (11). The first "block" row of the matrix $\mathbf{A}_2$ contains $\mathbf{A}_1$ and $\mathbf{B}_1$ only. We obtain each next row of this matrix by a cyclic shifting of the previous row one position to the right. We will call the $(n \times n)$-blocks $\mathbf{A}_1$ and $\mathbf{B}_1$ *generating blocks* or *generating matrices* of the matrix $\mathbf{A}_2$ and introduce the notation

$$\mathbf{A}_2 \sim \{\mathbf{A}_1, \mathbf{B}_1\}.$$

Then the eigenvalue problem takes the form

$$\begin{pmatrix} \mathbf{A}_1 & \mathbf{B}_1 & 0 & 0 & \cdots & \mathbf{B}_1 \\ \mathbf{B}_1 & \mathbf{A}_1 & \mathbf{B}_1 & 0 & \cdots & 0 \\ 0 & \mathbf{B}_1 & \mathbf{A}_1 & \mathbf{B}_1 & \cdots & 0 \\ \cdots & \cdots & \cdots & \cdots & \cdots & \cdots \\ \mathbf{B}_1 & 0 & \cdots & 0 & \mathbf{B}_1 & \mathbf{A}_1 \end{pmatrix} \begin{pmatrix} \mathbf{x}^{(1)} \\ \mathbf{x}^{(2)} \\ \mathbf{x}^{(3)} \\ \vdots \\ \mathbf{x}^{(n)} \end{pmatrix} = \mu \begin{pmatrix} \mathbf{x}^{(1)} \\ \mathbf{x}^{(2)} \\ \mathbf{x}^{(3)} \\ \vdots \\ \mathbf{x}^{(n)} \end{pmatrix}, \text{ where } \mathbf{x}^{(i)} = \begin{pmatrix} x_1^{(i)} \\ x_2^{(i)} \\ x_3^{(i)} \\ \vdots \\ x_n^{(i)} \end{pmatrix}, \ i = 1, 2, ..n. \qquad (13)$$

Here $\mathbf{x}^{(1)}, \ldots, \mathbf{x}^{(n)}$ are the vector-components of the $n$-dimensional eigenvector of the larger dimension $n^2$.

We present the space $\mathbf{R}^{n^2}$ as an orthogonal sum of $n$ subspaces $\mathbf{R}^n$ on which the generating matrices $\mathbf{A}_1$ and $\mathbf{B}_1$ are defined: $\mathbf{R}^{n^2} = \mathbf{R}^n \oplus \mathbf{R}^n \oplus ... \oplus \mathbf{R}^n$. In each of the subspaces $\mathbf{R}^n$ we change from the Cartesian unit vectors to a basis formed by $n$ eigenvectors (4) of the matrix $\mathbf{J}(1)$. Since, in this basis the generating matrices $\mathbf{A}_1$ and $\mathbf{B}_1$ are diagonal, we can rewrite the equations (13) as

$$\begin{pmatrix} w_1\lambda_1 & \cdots & 0 & w_1+w_2\lambda_1 & \cdots & 0 & w_1+w_2\lambda_1 & \cdots & 0 \\ \vdots & \ddots & \vdots & \vdots & \ddots & \vdots & \vdots & \ddots & \vdots \\ 0 & \cdots & w_1\lambda_n & 0 & \cdots & w_1+w_2\lambda_n & 0 & \cdots & w_1+w_2\lambda_n \\ w_1+w_2\lambda_1 & \cdots & 0 & w_1\lambda_1 & \cdots & 0 & & & \\ \vdots & \ddots & \vdots & \vdots & \ddots & \vdots & & 0 & \\ 0 & \cdots & 1+w_2\lambda_n & 0 & \cdots & w_1\lambda_n & & & \\ & \cdots & & & \cdots & & \ddots & \cdots & \\ w_1+w_2\lambda_1 & \cdots & 0 & & & & w_1\lambda_1 & \cdots & 0 \\ \vdots & \ddots & \vdots & & 0 & & \vdots & \ddots & \vdots \\ 0 & \cdots & w_1+w_2\lambda_n & & & & 0 & \cdots & w_1\lambda_n \end{pmatrix} \cdot \begin{pmatrix} x_1^{(1)} \\ \vdots \\ x_n^{(1)} \\ x_1^{(2)} \\ \vdots \\ x_n^{(2)} \\ \vdots \\ x_1^{(n)} \\ \vdots \\ x_n^{(n)} \end{pmatrix} = \mu \begin{pmatrix} x_1^{(1)} \\ \vdots \\ x_n^{(1)} \\ x_1^{(2)} \\ \vdots \\ x_n^{(2)} \\ \vdots \\ x_1^{(n)} \\ \vdots \\ x_n^{(n)} \end{pmatrix} \quad (14)$$

This $n^2$-dimensional system of equations splits into $n$ independent $n$-dimensional systems. Indeed, selecting the first, the $(n+1)$-th, the $(2n+1)$-th, ..., and the $(n^2-n+1)$-th rows of the system (14), we obtain the following system of equations:

$$\begin{array}{rcl} w_1\lambda_1 x_1^{(1)} + (w_1+w_2\lambda_1)x_1^{(2)} + 0 & \cdots & (w_1+w_2\lambda_1)x_1^{(n)} = \mu x_1^{(1)} \\ (w_1+w_2\lambda_1)x_1^{(1)} + w_1\lambda_1 x_1^{(2)} + (w_1+w_2\lambda_1)x_1^{(3)} & \cdots & 0 = \mu x_1^{(2)} \\ 0 + (w_1+w_2\lambda_1)x_1^{(2)} + w_1\lambda_1 x_1^{(3)} & \cdots & 0 = \mu x_1^{(3)} \\ \cdots & \cdots & \cdots \\ (w_1+w_2\lambda_1)x_1^{(1)} + 0 + 0 & \cdots & w_1\lambda_1 x_1^{(n)} = \mu x_1^{(n)} \end{array} \quad (15)$$

The variables $x_1^{(1)}$, $x_1^{(2)}$, ..., and $x_1^{(n)}$ of the system (15) do not enter any other equation of the system (14). Consequently, we can solve the system of equations (15) independently without account for other variables. In the same way, combining the second, the $(n+2)$-th, the $(2n+2)$-th, ..., and the $(n^2-n+2)$-th rows of the system (14) we obtain another system of equations analogous to the system (15), where we have to replace $\lambda_1$ by $\lambda_2$ and the set of variables $\{x_1^{(i)}\}_1^n$ by the set $\{x_2^{(i)}\}_1^n$. Proceeding as above, we see that the system (14) splits into $n$ systems of the same type

$$\begin{array}{rcl} w_1\lambda_i x_i^{(1)} + (w_1+w_2\lambda_i)x_i^{(2)} + 0 & \cdots & (w_1+w_2\lambda_i)x_i^{(n)} = \mu x_i^{(1)} \\ (w_1+w_2\lambda_i)x_i^{(1)} + w_1\lambda_i x_i^{(2)} + (w_1+w_2\lambda_i)x_i^{(3)} & \cdots & 0 = \mu x_i^{(2)} \\ 0 + (w_1+w_2\lambda_i)x_i^{(2)} + w_1\lambda_i x_i^{(3)} & \cdots & 0 = \mu x_i^{(3)}, \; i=1,2,...,n, \\ \cdots & \cdots & \cdots \\ (w_1+w_2\lambda_i)x_i^{(1)} + 0 + 0 & \cdots & w_1\lambda_i x_i^{(n)} = \mu x_i^{(n)} \end{array} \quad (16)$$

which have to be solved independently.

Let us simplify the system (16) by moving the diagonal terms of each equation to the right-hand side and by dividing all the coefficients by the same factor $w_1+w_2\lambda_i$. This way we obtain the following system of equations that coincides with the system (2):

$$\begin{array}{rcl} 0 + x_i^{(2)} + 0 \ldots + x_i^{(n)} & = & v_i \cdot x_i^{(1)} \\ x_i^{(1)} + 0 + x_i^{(3)} \ldots + 0 & = & v_i \cdot x_i^{(2)} \\ .. & .. & .. \\ x_i^{(1)} + 0 + 0 \ldots x_i^{(n-1)} + 0 & = & v_i \cdot x_i^{(n)} \end{array}$$, where $v_i = \dfrac{\mu - w_1\lambda_i}{w_1+w_2\lambda_i}$.

Consequently, according to Eq. (3) the variable $v_i$ takes $n$ different values $\{\lambda_j\}_{j=1}^n$:

$$v_i = \lambda_j, \; j=1,2,...,n \;\Rightarrow\; \mu = (w_1+w_2\lambda_i)\cdot\lambda_j + w_1\lambda_i, \text{ where } j=1,2,...,n.$$

Varying the index $i$ from 1 to $n$, we finally obtain

$$\mu_{ij} = w_1 \cdot (\lambda_i + \lambda_j) + w_2 \cdot \lambda_i \lambda_j, \quad i, j = 1, 2, \ldots n. \tag{17}$$

In the case of periodic boundary conditions, this equation defines the eigenvalues of the two-dimensional spin Hamiltonian when we account for the interactions with the spins of the two nearest coordination spheres only. It is easy to show that the eigenvectors are the Kronecker products of the eigenvectors $\mathbf{f}_i$ and $\mathbf{f}_j$ (4):

$$\mathbf{F}_{ij} = \mathbf{f}_i \otimes \mathbf{f}_j = \left( f_i^{(1)} f_j^{(1)}, f_i^{(1)} f_j^{(2)}, \ldots, f_i^{(1)} f_j^{(n)}, f_i^{(2)} f_j^{(1)}, f_i^{(2)} f_j^{(2)}, \ldots, f_i^{(2)} f_j^{(n)}, \ldots, f_i^{(n)} f_j^{(1)}, f_i^{(n)} f_j^{(2)}, \ldots, f_i^{(n)} f_j^{(n)} \right)^+, \tag{18}$$

where the superscript "+" defines the transposition operation.

**2.** In the same manner, we can also examine the three-dimensional Ising model. Here we outline the key points of this analysis and present the main formulas only.

The total number of spins of the three-dimensional Ising system is equal to $n^3$, and by $\mathbf{A}_3$ we denote its $(n^3 \times n^3)$ Hamiltonian matrix. It is convenient to present this matrix in the form analogous to Eq. (12):

$$\mathbf{A}_3 = \begin{pmatrix} \mathbf{A}_2 & \mathbf{B}_2 & 0 & 0 & \cdots & \mathbf{B}_2 \\ \mathbf{B}_2 & \mathbf{A}_2 & \mathbf{B}_2 & 0 & \cdots & 0 \\ 0 & \mathbf{B}_2 & \mathbf{A}_2 & \mathbf{B}_2 & \cdots & 0 \\ \cdots & \cdots & \cdots & \cdots & \cdots & \cdots \\ \mathbf{B}_2 & 0 & \cdots & 0 & \mathbf{B}_2 & \mathbf{A}_2 \end{pmatrix}.$$

The matrix $\mathbf{A}_3$ consists of the generating blocks $\mathbf{A}_2$ and $\mathbf{B}_2$ of dimension $(n^2 \times n^2)$: $\mathbf{A}_3 \sim (\mathbf{A}_2, \mathbf{B}_2)$. Equation (12) defines the matrix $\mathbf{A}_2$; its generating blocks of dimension $(n \times n)$ are $\mathbf{A}_1 = w_1 \mathbf{J}(1)$ and $\mathbf{B}_1 = w_1 \mathbf{I} + w_2 \mathbf{J}(1)$ (see Eq. (11)). It is easy to show that the $(n^2 \times n^2)$-matrix $\mathbf{B}_2$ is

$$\mathbf{B}_2 = \begin{pmatrix} \mathbf{B}_1 & \mathbf{D}_1 & 0 & 0 & \cdots & \mathbf{D}_1 \\ \mathbf{D}_1 & \mathbf{B}_1 & \mathbf{D}_1 & 0 & \cdots & 0 \\ 0 & \mathbf{D}_1 & \mathbf{B}_1 & \mathbf{D}_1 & \cdots & 0 \\ \cdots & \cdots & \cdots & \cdots & \cdots & \cdots \\ \mathbf{D}_1 & 0 & \cdots & 0 & \mathbf{D}_1 & \mathbf{B}_1 \end{pmatrix}, \text{ where } \mathbf{D}_1 = w_2 \mathbf{I} + w_3 \mathbf{J}(1). \tag{19}$$

The generating blocks of this matrix are $\mathbf{B}_1$ and $\mathbf{D}_1$: $\mathbf{B}_2 \sim (\mathbf{B}_1, \mathbf{D}_1)$.

By $\mu_{ij}^{(1)}$ we denote the eigenvalues of the matrix $\mathbf{B}_2$. We can easily calculate them if in all the equations of the previous Subsection we substitute the matrix $\mathbf{B}_2$ in place of $\mathbf{A}_2$. Then repeating the argumentation presented above we obtain

$$\mu_{ij}^{(1)} = w_1 + w_2 (\lambda_i + \lambda_j) + w_3 \lambda_i \lambda_j, \quad i, j = 1, 2, \ldots n.$$

The eigenvectors of the matrix $\mathbf{B}_2$ are the same Kronecker products $\mathbf{F}_{ij} = \mathbf{f}_i \otimes \mathbf{f}_j$ (18).

Let us present the space $\mathbf{R}^{n^3}$ as an orthogonal sum of the subspaces $\mathbf{R}^{n^2}$, where the matrices $\mathbf{A}_2$ and $\mathbf{B}_2$ are defined: $\mathbf{R}^{n^3} = \mathbf{R}^{n^2} \oplus \mathbf{R}^{n^2} \oplus \ldots \oplus \mathbf{R}^{n^2}$. In the subspaces $\mathbf{R}^{n^2}$ we change from the Cartesian basis to the basis consisting of the eigenvectors $\mathbf{F}_{ij}$ (18). In this basis, both generating $(n^2 \times n^2)$-blocks $\mathbf{A}_2$ and $\mathbf{B}_2$ are diagonal. Then the system of equations for the eigenvalues of the matrix $\mathbf{A}_3$ has a quasi-diagonal form analogous to the system (14):

$$\begin{pmatrix}
\mu_{11} & \cdots & 0 & \mu_{11}^{(1)} & \cdots & 0 & & & \mu_{11}^{(1)} & \cdots & 0 \\
\vdots & \ddots & \vdots & \vdots & \ddots & \vdots & & 0 & \cdots & \vdots & \ddots & \vdots \\
0 & \cdots & \mu_{nn} & 0 & \cdots & \mu_{nn}^{(1)} & & & 0 & \cdots & \mu_{nn}^{(1)} \\
\mu_{11}^{(1)} & \cdots & 0 & \mu_{11} & \cdots & 0 & \mu_{11}^{(1)} & \cdots & 0 \\
\vdots & \ddots & \vdots & \vdots & \ddots & \vdots & \vdots & \ddots & \vdots & \cdots & & 0 \\
0 & \cdots & \mu_{nn}^{(1)} & 0 & \cdots & \mu_{nn} & 0 & \cdots & \mu_{nn}^{(1)} \\
& \cdots & & & \cdots & & & \ddots & & & \cdots & \\
\mu_{11}^{(1)} & \cdots & 0 & & & & & & \mu_{11} & \cdots & 0 \\
\vdots & \ddots & \vdots & & 0 & & & 0 & \cdots & \vdots & \ddots & \vdots \\
0 & \cdots & \mu_{nn}^{(1)} & & & & & & 0 & \cdots & \mu_{nn}
\end{pmatrix}
\begin{pmatrix} x_{11}^{(1)} \\ \vdots \\ x_{nn}^{(1)} \\ x_{11}^{(2)} \\ \vdots \\ x_{nn}^{(2)} \\ \vdots \\ x_{11}^{(n)} \\ \vdots \\ x_{nn}^{(n)} \end{pmatrix}
= \mu \begin{pmatrix} x_{11}^{(1)} \\ \vdots \\ x_{nn}^{(1)} \\ x_{11}^{(2)} \\ \vdots \\ x_{nn}^{(2)} \\ \vdots \\ x_{11}^{(n)} \\ \vdots \\ x_{nn}^{(n)} \end{pmatrix}. \quad (20)$$

This system of equations splits into $n^2$ independent $n$-dimensional systems of the type (16):

$$\begin{array}{ccccccccl}
\mu_{ij} x_{ij}^{(1)} & + & \mu_{ij}^{(1)} x_{ij}^{(2)} & + & 0 & \cdots & \mu_{ij}^{(1)} x_{ij}^{(n)} & = & \mu x_{ij}^{(1)} \\
\mu_{ij}^{(1)} x_{ij}^{(1)} & + & \mu_{ij} x_{ij}^{(2)} & + & \mu_{ij}^{(1)} x_{ij}^{(3)} & \cdots & 0 & = & \mu x_{ij}^{(2)} \\
0 & + & \mu_{ij}^{(1)} x_{ij}^{(2)} & + & \mu_{ij} x_{ij}^{(3)} & \cdots & 0 & = & \mu x_{ij}^{(3)} \text{, where } i,j = 1,2,..,n. \\
\cdots & & \cdots & & \cdots & \cdots & \cdots & & \cdots \\
\mu_{ij}^{(1)} x_{ij}^{(1)} & + & 0 & + & 0 & \cdots & \mu_{ij} x_{ij}^{(n)} & = & \mu x_{ij}^{(n)}
\end{array}$$

We transform these systems of equations just as we have done when solving the systems (16). Then we obtain the eigenvalues of the three-dimensional Ising Hamiltonian in question:

$$\mu_{ijk} = \mu_{ij} + \lambda_k \cdot \mu_{ij}^{(1)} = w_1 \cdot (\lambda_i + \lambda_j + \lambda_k) + w_2 \cdot (\lambda_i \lambda_j + \lambda_j \lambda_k + \lambda_k \lambda_i) + w_3 \cdot \lambda_i \lambda_j \lambda_k \text{, where } i,j,k = 1,2,...n. \quad (21)$$

Here the eigenvectors are the Kronecker products of the eigenvectors of the one-dimensional Ising Hamiltonian:

$$\mathbf{F}_{ijk} = \mathbf{F}_{ij} \otimes \mathbf{f}_k = \mathbf{f}_i \otimes \mathbf{f}_j \otimes \mathbf{f}_k.$$

### III. 2D- and 3D-Ising systems with free boundaries

**1.** All the arguments of Subsection 1 of the previous Section also hold when we examine the case of free boundaries. All we need is to replace the matrix $\mathbf{J}(1)$ by the matrix $\tilde{\mathbf{J}}(1)$, the eigenvalues $\lambda_i \equiv \lambda_i(1)$ (3) by the eigenvalues $\tilde{\lambda}_i \equiv \tilde{\lambda}_i(1)$ (9), and the eigenvectors $\mathbf{f}_i$ (4) by the eigenvectors $\tilde{\mathbf{f}}_i$ (9). Then repeating all the calculations we obtain an analog of the formula (17):

$$\tilde{\mu}_{ij} = w_1 \cdot (\tilde{\lambda}_j + \tilde{\lambda}_i) + w_2 \cdot \tilde{\lambda}_i \tilde{\lambda}_j, \quad i,j = 1,2,...n. \quad (22)$$

Now we can easily explain why for the planar free boundary Ising system it is not possible to generalize Eq. (22) to the case of a larger number of the neighbors.

In Subsection 1 of the previous Section, the key point was the change from the Cartesian basis to the basis consisting of the eigenvectors $\{\mathbf{f}^{(i)}\}_1^n$ (4), where the matrices $\mathbf{A}_1$ and $\mathbf{B}_1$ were diagonal and the system of equations (14) had the quasi-diagonal form. Let us emphasize that in the case of free boundary conditions for the same purpose we have to use the basis $\{\tilde{\mathbf{f}}^{(i)}\}_1^n$ (9) of the eigenvectors of the matrix $\tilde{\mathbf{J}}(1)$.

Note, only in this rather simple case, the interaction matrix $\mathbf{A}_2$ consists of two generating blocks (11) and, in addition, their structures are very simple. In the general case, the number of the generating blocks is larger and their structures are more complex. For example, if in the planar Ising system we account for the interactions with the spins from the third coordination sphere, the Hamiltonian $\mathbf{A}_2$ becomes more complicated (see also Subsection 1 of the next Section):

$$\mathbf{A}_2 \sim (\mathbf{A}_1, \mathbf{B}_1, \mathbf{C}_1), \text{ where } \mathbf{A}_1 = w_1 \mathbf{J}(1) + w_3 \mathbf{J}(2), \ \mathbf{B}_1 = w_1 \mathbf{I} + w_2 \mathbf{J}(1), \ \mathbf{C}_1 = w_3 \mathbf{I}. \tag{23}$$

In the case of periodic boundary conditions, we can diagonalize all these matrices simultaneously in the basis of the eigenvectors (4). As a result, we can split the large system (14) into $n$ independent systems (16). All other results are the consequence of this possibility.

On the contrary, in the case of free boundary conditions there is no basis in which we can diagonalize the matrices $\tilde{\mathbf{J}}(1)$ and $\tilde{\mathbf{J}}(2)$ simultaneously. Consequently, it is impossible to diagonalize simultaneously all the three generating blocks $\tilde{\mathbf{A}}_1$, $\tilde{\mathbf{B}}_1$, and $\tilde{\mathbf{C}}_1$ (23). This means that after Eq. (14) all the equations become invalid.

In the case of the two-dimensional Ising system with free boundaries, the expressions (22) are correct, but their generalization to the case of a larger number of the neighbors is impossible.

**2.** In the same manner, we can analyze the three-dimensional Ising systems. If, similarly to Subsection 2 of the previous Section, we account only for the interactions with the spins of the first three coordination spheres, all the arguments of this Subsection remain valid. In this case, the generating blocks of the first level are the $(n \times n)$-matrices

$$\tilde{\mathbf{A}}_1 = w_1 \tilde{\mathbf{J}}(1), \ \tilde{\mathbf{B}}_1 = w_1 \mathbf{I} + w_2 \tilde{\mathbf{J}}(1), \text{ and } \tilde{\mathbf{D}}_1 = w_2 \mathbf{I} + w_3 \tilde{\mathbf{J}}(1)$$

(see Eqs. (11) and (19)). We can diagonalize these three generating blocks simultaneously in the basis of the vectors $\{\tilde{\mathbf{f}}_i\}_1^n$ (9). Then the $(n^2 \times n^2)$-matrices $\tilde{\mathbf{A}}_2$ and $\tilde{\mathbf{B}}_2$ (see equations (12) and (19), respectively) are diagonal in the basis of the Kronecker products $\tilde{\mathbf{F}}_{ij} = \tilde{\mathbf{f}}_i \otimes \tilde{\mathbf{f}}_j$ and the $(n^3 \times n^3)$-eigenvalue problem has a quasi-diagonal form (20). After adding tilde where it is necessary, from Eq. (21) we obtain the expressions for the eigenvalues $\tilde{\mu}_{ijk}$:

$$\tilde{\mu}_{ijk} = w_1 \cdot (\tilde{\lambda}_i + \tilde{\lambda}_j + \tilde{\lambda}_k) + w_2 \cdot (\tilde{\lambda}_i \tilde{\lambda}_j + \tilde{\lambda}_j \tilde{\lambda}_k + \tilde{\lambda}_k \tilde{\lambda}_i) + w_3 \cdot \tilde{\lambda}_i \tilde{\lambda}_j \tilde{\lambda}_k, \ i, j, k = 1, 2, \ldots n. \tag{24}$$

The situation is different when, for the three-dimensional Ising system, we attempt to account for more spins (for example, for the spins of the fourth coordination sphere). In this case, there are more generating blocks of all levels and their structures are more complex (see also Subsection 2 of the next Section). For example, the generating blocks of the first level has the form:

$$\tilde{\mathbf{A}}_1 = w_1 \tilde{\mathbf{J}}(1) + w_4 \tilde{\mathbf{J}}(2), \ \tilde{\mathbf{B}}_1 = w_1 \mathbf{I} + w_2 \tilde{\mathbf{J}}(1), \ \tilde{\mathbf{C}}_1 = w_4 \mathbf{I}, \text{ and } \tilde{\mathbf{D}}_1 = w_2 \mathbf{I} + w_3 \tilde{\mathbf{J}}(1).$$

Due to the presence of the matrix $\tilde{\mathbf{J}}(2)$ in the expression for the matrix $\tilde{\mathbf{A}}_1$, there is no basis in which we can diagonalize all these matrices simultaneously. Consequently, the Kronecker products $\tilde{\mathbf{F}}_{ij} = \tilde{\mathbf{f}}_i \otimes \tilde{\mathbf{f}}_j$ are not the eigenvectors of the $(n^2 \times n^2)$-matrix $\tilde{\mathbf{A}}_2$ and all the arguments of Subsection 2 of previous Section are not valid.

For the three-dimensional Ising system with free boundaries, equations (24) are valid, but their generalization to the case of a larger number of neighbors is impossible. This concludes our examination of the Ising system with free boundaries. The remaining part of the paper concerns the systems with periodic boundary conditions.

### V. 2D and 3D Ising models with periodic boundary conditions: general case

In this Section, we develop another method for calculation of the eigenvalues of the multidimensional Ising Hamiltonian. Computationally, it is more efficient. The method presents the eigenvectors of the multidimensional problem as the Kronecker products of the eigenvectors of the one-dimensional Ising Hamiltonian (4). For periodic boundary conditions, this procedure is always correct.

A planar lattice is a sublattice of a cubic lattice. Consequently, sometimes the distances between the pairs of the nodes of these two lattices coincide, but the numbers of the nodes differ. To avoid misunderstanding, when in the three-dimensional case we will use the formulas relevant to planar systems, we will supply the interaction constants with an extra superscript indicating the dimension of the lattice. Thus, $w_k^{(2)}$ will be the constant of interaction with the spins of the $k$-th coordination sphere of the planar lattice; $w_k^{(3)}$ will define the interaction with the spins of $k$-th coordination sphere of the cubic lattice.

**1. Two-dimensional Ising model with five neighbors.** The first five minimal distances between the nodes of the two-dimensional square lattice are equal to $l_1 = 1$, $l_2 = \sqrt{2}$, $l_3 = 2$, $l_4 = \sqrt{5}$, and $l_5 = 2\sqrt{2}$. In Fig. 1, the circles mark the lattice nodes, and the number of the node corresponds to one of the listed distances between this node and the given node (the square in the lower left node).

Let us write the Hamiltonian as a block matrix

$$\mathbf{A}_2 = \begin{pmatrix} \mathbf{A}_1 & \mathbf{B}_1 & \mathbf{C}_1 & 0 & \ldots & \mathbf{C}_1 & \mathbf{B}_1 \\ \mathbf{B}_1 & \mathbf{A}_1 & \mathbf{B}_1 & \mathbf{C}_1 & \ldots & 0 & \mathbf{C}_1 \\ \mathbf{C}_1 & \mathbf{B}_1 & \mathbf{A}_1 & \mathbf{B}_1 & \ldots & 0 & 0 \\ 0 & \mathbf{C}_1 & \mathbf{B}_1 & \mathbf{A}_1 & \ldots & 0 & 0 \\ \ldots & \ldots & \ldots & \ldots & \ldots & \ldots & \ldots \\ \mathbf{C}_1 & 0 & 0 & 0 & \ldots & \mathbf{A}_1 & \mathbf{B}_1 \\ \mathbf{B}_1 & \mathbf{C}_1 & 0 & 0 & \ldots & \mathbf{B}_1 & \mathbf{A}_1 \end{pmatrix}, \qquad (25)$$

where the $(n \times n)$ matrices $\mathbf{A}_1$, $\mathbf{B}_1$, and $\mathbf{C}_1$ are equal to

$$\begin{aligned}
\mathbf{A}_1 &= w_1^{(2)} \mathbf{J}(1) + w_3^{(2)} \mathbf{J}(2), \\
\mathbf{B}_1 &= w_1^{(2)} \mathbf{I} + w_2^{(2)} \mathbf{J}(1) + w_4^{(2)} \mathbf{J}(2), \\
\mathbf{C}_1 &= w_3^{(2)} \mathbf{I} + w_4^{(2)} \mathbf{J}(1) + w_5^{(2)} \mathbf{J}(2).
\end{aligned} \qquad (26)$$

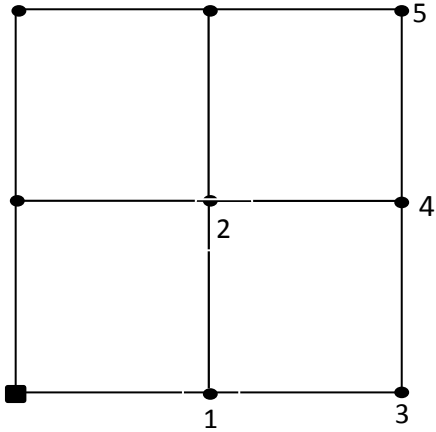
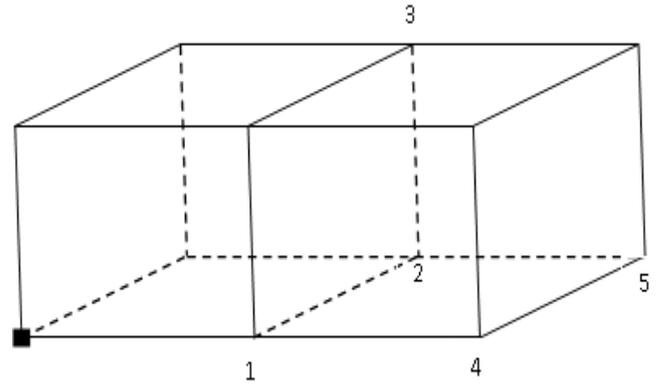

**Fig. 1.** Square lattice: first five neighbors of given node.

**Fig. 2.** Cubic lattice: first five neighbors of given node.

If $\mathbf{F}_2$ is an eigenvector of the matrix $\mathbf{A}_2$, the corresponding eigenvalue is equal to $\mu = \mathbf{F}_2^+ \mathbf{A}_2 \mathbf{F}_2$. We are looking for the vectors $\mathbf{F}_2$ in the form of the Kronecker products of the eigenvectors $\{\mathbf{f}_i\}_i^n$ (see Eq. (4)): $\mathbf{F}_2 = \mathbf{f}_r \otimes \mathbf{f}_k$. Then we obtain for the eigenvalue $\mu_{kr}$

$$\mu_{kr} = \mathbf{f}_k^+ \otimes \mathbf{f}_r^+ \begin{pmatrix} \mathbf{A}_1 & \mathbf{B}_1 & \mathbf{C}_1 & 0 & \ldots & \mathbf{C}_1 & \mathbf{B}_1 \\ \mathbf{B}_1 & \mathbf{A}_1 & \mathbf{B}_1 & \mathbf{C}_1 & \ldots & 0 & \mathbf{C}_1 \\ \mathbf{C}_1 & \mathbf{B}_1 & \mathbf{A}_1 & \mathbf{B}_1 & \ldots & 0 & 0 \\ 0 & \mathbf{C}_1 & \mathbf{B}_1 & \mathbf{A}_1 & \ldots & 0 & 0 \\ \ldots & \ldots & \ldots & \ldots & \ldots & \ldots & \ldots \\ \mathbf{C}_1 & 0 & 0 & 0 & \ldots & \mathbf{A}_1 & \mathbf{B}_1 \\ \mathbf{B}_1 & \mathbf{C}_1 & 0 & 0 & \ldots & \mathbf{B}_1 & \mathbf{A}_1 \end{pmatrix} \mathbf{f}_r \otimes \mathbf{f}_k .$$

We use the associativity of the matrix multiplication:

$$\mu_{kr} = \mathbf{f}_k^+ \begin{pmatrix} \mathbf{f}_r^+\mathbf{A}_1\mathbf{f}_r & \mathbf{f}_r^+\mathbf{B}_1\mathbf{f}_r & \mathbf{f}_r^+\mathbf{C}_1\mathbf{f}_r & 0 & \ldots & \mathbf{f}_r^+\mathbf{B}_1\mathbf{f}_r \\ \mathbf{f}_r^+\mathbf{B}_1\mathbf{f}_r & \mathbf{f}_r^+\mathbf{A}_1\mathbf{f}_r & \mathbf{f}_r^+\mathbf{B}_1\mathbf{f}_r & \mathbf{f}_r^+\mathbf{C}_1\mathbf{f}_r & \ldots & \mathbf{f}_r^+\mathbf{C}_1\mathbf{f}_r \\ \mathbf{f}_r^+\mathbf{C}_1\mathbf{f}_r & \mathbf{f}_r^+\mathbf{B}_1\mathbf{f}_r & \mathbf{f}_r^+\mathbf{A}_1\mathbf{f}_r & \mathbf{f}_r^+\mathbf{B}_1\mathbf{f}_r & \ldots & 0 \\ 0 & \mathbf{f}_r^+\mathbf{C}_1\mathbf{f}_r & \mathbf{f}_r^+\mathbf{B}_1\mathbf{f}_r & \mathbf{f}_r^+\mathbf{A}_1\mathbf{f}_r & \ldots & 0 \\ \ldots & \ldots & \ldots & \ldots & \ldots & \ldots \\ \mathbf{f}_r^+\mathbf{B}_1\mathbf{f}_r & \mathbf{f}_r^+\mathbf{C}_1\mathbf{f}_r & 0 & 0 & \ldots & \mathbf{f}_r^+\mathbf{A}_1\mathbf{f}_r \end{pmatrix} \mathbf{f}_k = \mathbf{f}_k^+ \cdot \mathbf{M} \cdot \mathbf{f}_k ,$$

Let $\mathbf{M}$ denote $(n \times n)$ matrix. It is easy to see that

$$\mathbf{M} = \left(\mathbf{f}_r^+\mathbf{A}_1\mathbf{f}_r\right) \cdot \mathbf{I} + \left(\mathbf{f}_r^+\mathbf{B}_1\mathbf{f}_r\right) \cdot \mathbf{J}(1) + \left(\mathbf{f}_r^+\mathbf{C}_1\mathbf{f}_r\right) \cdot \mathbf{J}(2), \tag{27}$$

where according to Eq. (26) we have

$$\begin{aligned}
\left(\mathbf{f}_r^+\mathbf{A}_1\mathbf{f}_r\right) &= w_1^{(2)}\lambda_r(1) + w_3^{(2)}\lambda_r(2), \\
\left(\mathbf{f}_r^+\mathbf{B}_1\mathbf{f}_r\right) &= w_1^{(2)} + w_2^{(2)}\lambda_r(1) + w_4^{(2)}\lambda_r(2), \\
\left(\mathbf{f}_r^+\mathbf{C}_1\mathbf{f}_r\right) &= w_3^{(2)} + w_4^{(2)}\lambda_r(1) + w_5^{(2)}\lambda_r(2).
\end{aligned} \tag{28}$$

Then $\mu_{kr}$ is equal to

$$\mu_{kr} = \left(\mathbf{f}_r^+\mathbf{A}_1\mathbf{f}_r\right) + \left(\mathbf{f}_r^+\mathbf{B}_1\mathbf{f}_r\right) \cdot \lambda_k(1) + \left(\mathbf{f}_r^+\mathbf{C}_1\mathbf{f}_r\right) \cdot \lambda_k(2).$$

Making use of Eq. (28) we obtain:

$$\mu_{kr} = w_1^{(2)}\left(\lambda_k + \lambda_r\right) + w_2^{(2)}\lambda_k\lambda_r + w_3^{(2)}\left(\lambda_k(2) + \lambda_r(2)\right) + w_4^{(2)}\left(\lambda_k\lambda_r(2) + \lambda_r\lambda_k(2)\right) + w_5^{(2)}\lambda_k(2)\lambda_r(2).$$

If now we account for the relation $\lambda_k(2) = \lambda_k^2 - 2$ then

$$\mu_{kr} = w_1^{(2)}\left(\lambda_k + \lambda_r\right) + w_2^{(2)}\lambda_k\lambda_r + w_3^{(2)}\left(\lambda_k^2 + \lambda_r^2 - 4\right) + w_4^{(2)}\left(\lambda_k\lambda_r^2 + \lambda_r\lambda_k^2 - 2(\lambda_k + \lambda_r)\right) + w_5^{(2)}\left(\lambda_k^2 - 2\right)\left(\lambda_r^2 - 2\right). \tag{29}$$

If in Eq. (29) we set $w_1^{(2)} = 1$ and $w_2^{(2)} = w_3^{(2)} = w_4^{(2)} = w_5^{(2)} = 0$, we obtain the known expressions (see [5]); if only the values of $w_3^{(2)}$, $w_4^{(2)}$, and $w_5^{(2)}$ are equal to zero, we obtain the expression (17). Comparing the equations (17) and (29) we see that accounting for additional more distant neighbors does not influence the result obtained previously for a smaller number of the neighbors. This is a consequence of the fact that all the matrices $\mathbf{J}(k)$ of the one-dimensional Ising model have the same set of the eigenvectors.

**2. Three-dimensional Ising model with five neighbors.** The first five minimal distances between the nodes of the three-dimensional cubic lattice are equal to $l_1 = 1$, $l_2 = \sqrt{2}$, $l_3 = \sqrt{3}$, $l_4 = 2,$ and $l_5 = \sqrt{5}$ (see Fig. 2). Let us compare the Figures 1 and 2. In the case of the two-dimensional lattice, the distance to the nearest neighbor $l_1$ coincides with the analogous distance for the three-dimensional lattice. Consequently, we should identify the interaction constants $w_1^{(2)}$ and $w_1^{(3)}$. Analyzing the figures it is easy to write down other pairs of the identical interaction constants:

$$w_1^{(2)} \equiv w_1^{(3)}, \ w_2^{(2)} \equiv w_2^{(3)}, \ w_3^{(2)} \equiv w_4^{(3)}, \ w_4^{(2)} \equiv w_5^{(3)}. \tag{30}$$

These identities allow us to use Eq. (26) obtained for the two-dimensional Ising model in the three-dimensional case. Since the fifth coordination sphere of the planar model is not among the first five coordination spheres of the cubic model, we have to set $w_5^{(2)} = 0$.

The Hamiltonian of the three-dimensional Ising system is defined by the matrix $\mathbf{A}_3$. It is convenient to present this matrix in the block form using the $(n^2 \times n^2)$ matrices $\mathbf{A}_2$, $\mathbf{B}_2$ и $\mathbf{C}_2$:

$$\mathbf{A}_3 = \begin{pmatrix} \mathbf{A}_2 & \mathbf{B}_2 & \mathbf{C}_2 & 0 & \dots & \mathbf{C}_2 & \mathbf{B}_2 \\ \mathbf{B}_2 & \mathbf{A}_2 & \mathbf{B}_2 & \mathbf{C}_2 & \dots & 0 & \mathbf{C}_2 \\ \mathbf{C}_2 & \mathbf{B}_2 & \mathbf{A}_2 & \mathbf{B}_2 & \dots & 0 & 0 \\ 0 & \mathbf{C}_2 & \mathbf{B}_2 & \mathbf{A}_2 & \dots & 0 & 0 \\ \dots & \dots & \dots & \dots & \dots & \dots & \dots \\ \mathbf{C}_2 & 0 & 0 & 0 & \dots & \mathbf{A}_2 & \mathbf{B}_2 \\ \mathbf{B}_2 & \mathbf{C}_2 & 0 & 0 & \dots & \mathbf{B}_2 & \mathbf{A}_2 \end{pmatrix}.$$

The block-matrix $\mathbf{A}_2$ has exactly the same form as the matrix (25); its generating $(n \times n)$ matrices differ from (26) only by the notations for the interaction constants (see Eq. (30)):

$$\mathbf{A}_1 = w_1^{(3)} \mathbf{J}(1) + w_4^{(3)} \mathbf{J}(2),$$
$$\mathbf{B}_1 = w_1^{(3)} \mathbf{I} + w_2^{(3)} \mathbf{J}(1) + w_5^{(3)} \mathbf{J}(2),$$
$$\mathbf{C}_1 = w_4^{(3)} \mathbf{I} + w_5^{(3)} \mathbf{J}(1).$$

The two remaining $(n^2 \times n^2)$ block-matrices are

$$\mathbf{B}_2 = \begin{pmatrix} \mathbf{B}_1 & \mathbf{D}_1 & \mathbf{E}_1 & 0 & \cdots & \mathbf{E}_1 & \mathbf{D}_1 \\ \mathbf{D}_1 & \mathbf{B}_1 & \mathbf{D}_1 & \mathbf{E}_1 & & 0 & \mathbf{E}_1 \\ \mathbf{E}_1 & \mathbf{D}_1 & \mathbf{B}_1 & \mathbf{D}_1 & & 0 & 0 \\ \dots & \dots & \dots & \dots & \dots & \dots & \dots \\ \mathbf{D}_1 & \mathbf{E}_1 & 0 & 0 & \cdots & \mathbf{D}_1 & \mathbf{B}_1 \end{pmatrix}, \quad \mathbf{C}_2 = \begin{pmatrix} \mathbf{C}_1 & \mathbf{E}_1 & 0 & \cdots & \mathbf{E}_1 \\ \mathbf{E}_1 & \mathbf{C}_1 & \mathbf{E}_1 & \cdots & 0 \\ 0 & \mathbf{E}_1 & \mathbf{C}_1 & \cdots & 0 \\ \dots & \dots & \dots & \dots & \dots \\ \mathbf{E}_1 & 0 & 0 & \cdots & \mathbf{C}_1 \end{pmatrix}, \text{ where } \begin{aligned} \mathbf{D}_1 &= w_2^{(3)} \mathbf{I} + w_3^{(3)} \mathbf{J}(1), \\ \mathbf{E}_1 &= w_5^{(3)} \mathbf{I}. \end{aligned}$$

We seek the eigenvectors $\mathbf{F}_3$ of the matrix $\mathbf{A}_3$ as the Kronecker products of the eigenvectors of the one- and two-dimensional models:

$$\mathbf{F}_3 = \mathbf{F}_2 \otimes \mathbf{f}_m, \text{ where } \mathbf{F}_2 = \mathbf{f}_r \otimes \mathbf{f}_k, \quad k, r, m = 1, 2, ..n.$$

Then the eigenvalue $\mu_{krm} = \mathbf{F}_3^+ \mathbf{A}_3 \mathbf{F}_3$ is

$$\mu_{krm} = \mathbf{f}_m^+ \begin{pmatrix} \mathbf{F}_2^+ \mathbf{A}_2 \mathbf{F}_2 & \mathbf{F}_2^+ \mathbf{B}_2 \mathbf{F}_2 & \mathbf{F}_2^+ \mathbf{C}_2 \mathbf{F}_2 & 0 & \dots & \mathbf{F}_2^+ \mathbf{C}_2 \mathbf{F}_2 & \mathbf{F}_2^+ \mathbf{B}_2 \mathbf{F}_2 \\ \mathbf{F}_2^+ \mathbf{B}_2 \mathbf{F}_2 & \mathbf{F}_2^+ \mathbf{A}_2 \mathbf{F}_2 & \mathbf{F}_2^+ \mathbf{B}_2 \mathbf{F}_2 & \mathbf{F}_2^+ \mathbf{C}_2 \mathbf{F}_2 & \dots & 0 & \mathbf{F}_2^+ \mathbf{C}_2 \mathbf{F}_2 \\ \mathbf{F}_2^+ \mathbf{C}_2 \mathbf{F}_2 & \mathbf{F}_2^+ \mathbf{B}_2 \mathbf{F}_2 & \mathbf{F}_2^+ \mathbf{A}_2 \mathbf{F}_2 & \mathbf{F}_2^+ \mathbf{B}_2 \mathbf{F}_2 & \dots & 0 & 0 \\ 0 & \mathbf{F}_2^+ \mathbf{C}_2 \mathbf{F}_2 & \mathbf{F}_2^+ \mathbf{B}_2 \mathbf{F}_2 & \mathbf{F}_2^+ \mathbf{A}_2 \mathbf{F}_2 & \dots & 0 & 0 \\ \dots & \dots & \dots & \dots & \dots & \dots & \dots \\ \mathbf{F}_2^+ \mathbf{C}_2 \mathbf{F}_2 & 0 & 0 & 0 & \dots & \mathbf{F}_2^+ \mathbf{A}_2 \mathbf{F}_2 & \mathbf{F}_2^+ \mathbf{B}_2 \mathbf{F}_2 \\ \mathbf{F}_2^+ \mathbf{B}_2 \mathbf{F}_2 & \mathbf{F}_2^+ \mathbf{C}_2 \mathbf{F}_2 & 0 & 0 & \dots & \mathbf{F}_2^+ \mathbf{B}_2 \mathbf{F}_2 & \mathbf{F}_2^+ \mathbf{A}_2 \mathbf{F}_2 \end{pmatrix} \mathbf{f}_m.$$

When in this equation we write the $(n \times n)$-matrix in the form analogous to Eq. (27), we obtain:

$$\mu_{krm} = (\mathbf{F}_2^+ \mathbf{A}_2 \mathbf{F}_2) + (\mathbf{F}_2^+ \mathbf{B}_2 \mathbf{F}_2) \cdot \lambda_m(1) + (\mathbf{F}_2^+ \mathbf{C}_2 \mathbf{F}_2) \cdot \lambda_m(2). \quad (31)$$

To calculate $(\mathbf{F}_2^+ \mathbf{A}_2 \mathbf{F}_2)$ we use Eq. (29) derived previously, but we have to adjust the weights taking into account Eq. (30):

$$(\mathbf{F}_{kr}^+ \mathbf{A}_2 \mathbf{F}_{kr}) = w_1^{(3)} (\lambda_k + \lambda_r) + w_2^{(3)} \lambda_k \lambda_r + w_4^{(3)} (\lambda_k^2 + \lambda_r^2 - 4) + w_5^{(3)} (\lambda_k \lambda_r^2 + \lambda_r \lambda_k^2 - 2(\lambda_k + \lambda_r)).$$

We calculate the values $(\mathbf{F}_2^+ \mathbf{B}_2 \mathbf{F}_2)$ and $(\mathbf{F}_2^+ \mathbf{C}_2 \mathbf{F}_2)$ in the same way as we calculated $(\mathbf{F}_2^+ \mathbf{A}_2 \mathbf{F}_2)$ in the previous subsection. As a result, we obtain:

$$(\mathbf{F}_{kr}^+ \mathbf{B}_2 \mathbf{F}_{kr}) = w_5^{(3)} + w_2^{(3)} (\lambda_k + \lambda_r) + w_3^{(3)} \lambda_k \lambda_r + w_5^{(3)} (\lambda_k^2 + \lambda_r^2 - 4), \quad (\mathbf{F}_{kr}^+ \mathbf{C}_2 \mathbf{F}_{kr}) = w_4^{(3)} + w_5^{(3)} (\lambda_k + \lambda_r).$$

Finally, substituting the last three expressions in Eq. (31) we find:

$$\mu_{krm} = w_1^{(3)}(\lambda_k + \lambda_r + \lambda_m) + w_2^{(3)}(\lambda_k\lambda_r + \lambda_r\lambda_m + \lambda_m\lambda_k) + w_3^{(3)}\lambda_k\lambda_r\lambda_m + w_4^{(3)}\left[\lambda_k^2 + \lambda_r^2 + \lambda_m^2 - 6\right] +$$
$$+ w_5^{(3)}\left[(\lambda_k + \lambda_r)\lambda_m^2 + (\lambda_k + \lambda_m)\lambda_r^2 + (\lambda_r + \lambda_m)\lambda_k^2 - 4(\lambda_k + \lambda_r + \lambda_m)\right], \quad k, r, m = 1,..,n. \quad (32)$$

## VI. Discussion and conclusions

We obtained the analytical expressions for the eigenvalues of the 2D- and 3D-Ising Hamiltonians taking into account the interactions with the spins of several coordination spheres. We expressed the eigenvalues of the multidimensional Hamiltonian in terms of the interaction constants $\{w_r\}_1^k$ and the eigenvalues $\{\lambda_i\}_1^n$ of the one-dimensional Ising model with the interactions of the nearest neighbors. The expressions for the eigenvalues $\{\lambda_i\}_1^n$ are well known for periodic boundary conditions (3), as well as for free boundary conditions (9). However, even for the one-dimensional model these two types of boundary conditions lead to substantially different results. Such difference extends to the multidimensional Ising problem.

We begin with summarizing our conclusions for periodic boundary conditions. We recall that $\mathbf{J}(r)$ is the connection matrix in the one-dimensional model that accounts for the interaction with the spins of the $r$-th coordination sphere *only* (see Section II). All the matrices $\mathbf{J}(r)$ have the same set of the eigenvectors and their eigenvalues are equal to $\lambda_i(r) = P_r(\lambda_i)$, where $P_r(x)$ is the $r$-th degree polynomial. Then $\mathbf{A}_1 = w_1\mathbf{J}(1) + ... + w_k\mathbf{J}(k)$ is the Hamiltonian of the one-dimensional Ising system, where we account for the interactions with $k$ closest coordination spheres and the eigenvalues $\lambda_i(\mathbf{A}_1)$ are the weighted sums of the eigenvalues $\lambda_i(r)$:

$$\lambda_i(\mathbf{A}_1) = \sum_{r=1}^k w_r\lambda_i(r), \quad i = 1, 2, ..., n. \quad (33)$$

This property of the one-dimensional model allows us to account for interactions with the spins of an arbitrary number of the coordination spheres and to express the eigenvalues of the multidimensional Hamiltonian in terms of $\{\lambda_i\}_1^n$ and $\{w_r\}_1^k$ (see Eqs. (29) and (32)). It is important that when we take into account interactions with additional neighbors, new terms appear in the expression for the eigenvalues, but the obtained previously terms do not change.

Assuming a certain dependence of the interaction constant on the distance, with the aid of Eq. (33) we can obtain an asymptotic expression for the spectral density of the one-dimensional Hamiltonian. Then the obtained result can be extended to the cases of two- and three-dimensional systems. We plan to implement such procedure in what follows.

In the case of the free boundary systems, the situation is quite different even in the one-dimensional case. The reason is that the matrices $\tilde{\mathbf{J}}(r)$ do not commute. All these matrices have different sets of the eigenvectors and even for the simplest linear combination of these matrices, $\tilde{\mathbf{A}}_1 = w_1\tilde{\mathbf{J}}(1) + w_2\tilde{\mathbf{J}}(2) + ... + w_k\tilde{\mathbf{J}}(k)$ we cannot express the eigenvalues of the matrix $\tilde{\mathbf{A}}_1$ in terms of the eigenvalues of the matrices $\tilde{\mathbf{J}}(r)$. The same is also true in the multidimensional case with two exceptions (see Section IV). We can express the eigenvalues of the two-dimensional Ising Hamiltonian in terms of the eigenvalues $\tilde{\lambda}_i$ when we account for two closest coordination spheres only (see Eq. (19)). To express the eigenvalues of the three-dimensional Ising Hamiltonian in terms of $w_1$, $w_2$, $w_3$, and $\tilde{\lambda}_i$, we can account for not more than three closest coordination spheres (see Eq. (24)).

All the obtained results can be generalized to the case of the Ising model on a hypercube of an arbitrary dimension $d > 3$. The difference between periodic and free boundary conditions also remains.

This work was supported by State Program of Scientific Research Institute for System Analysis, Russian Academy of Sciences, project no. 0065-2019-0003.

The authors are grateful to Dr. Inna Kaganova and Prof. Marina Litinskaya for helpful discussions and their help when preparing this paper.